\documentstyle[12pt]{article}

\newcommand{\beq}{\begin{equation}}
\newcommand{\eeq}{\end{equation}}
\newcommand{\beqa}{\begin{eqnarray}}
\newcommand{\eeqa}{\end{eqnarray}}
\newcommand{\beqar}{\begin{eqnarray*}}
\newcommand{\eeqar}{\end{eqnarray*}}

\newcommand{\eps}{\epsilon}
\newcommand{\ga}{\gamma}
\newcommand{\Ga}{\Gamma}
\newcommand{\ka}{\kappa}

\renewcommand{\l}{\lambda}

\newcommand{\sig}{\sigma}

\newcommand{\z}{\zeta}

\newcommand{\eg}{{\it e.g.,}\ }
\newcommand{\ie}{{\it i.e.,}\ }
\newcommand{\labell}[1]{\label{#1}} %{\label{#1}\qquad_{#1}} %
\newcommand{\reef}[1]{(\ref{#1})}
\newcommand\prt{\partial}
\newcommand\veps{\varepsilon}
\newcommand\ls{\ell_s}

\newcommand\cL{{\cal L}}
\newcommand\tL{\tilde{\cal L}}
\newcommand\tG{{\widetilde G}}
\newcommand\tg{{\tilde g}}
\newcommand\tB{{\widetilde B}}
\newcommand\tC{{\widetilde C}}
\newcommand\Tr{{\rm Tr}}
\newcommand\M[2]{M^{#1}{}_{#2}}
\begin{document}

\thispagestyle{empty}
\rightline{\small hep-th/9809100 \hfill UCSB NSF-ITP-98-084}
\rightline{\small \hfill McGill/98-21}
\rightline{\small \hfill IPM/P-98/12}
\vspace*{1cm}

\begin{center}
{\bf \Large World-Volume Interactions\\[.25em]
on D-Branes}
\vspace*{1cm}

{Mohammad R. Garousi\footnote{E-mail:
mohammad@physics.ipm.ac.ir}}\\
\vspace*{0.2cm}
{\it Institute for Studies in Theoretical Physics and Mathematics IPM} \\
{P.O. Box 19395-5746, Tehran, Iran}\\
\vspace*{0.1cm}
and\\
\vspace*{0.1cm}
{\it Department of Physics, Birjand University, Birjand, Iran}\\
\vspace*{0.4cm}

Robert C. Myers\footnote{E-mail: rcm@itp.ucsb.edu}\\
\vspace*{0.2cm}
{\it Department of Physics, McGill University}\\
{\it Montr\'eal, PQ, H3A 2T8, Canada}\\
\vspace*{0.1cm}
and\\
\vspace*{0.1cm}
{\it Institute of Theoretical Physics,
University of California}\\
{\it Santa Barbara, CA 93106-4030, USA}\\
\vspace{2cm}
ABSTRACT
\end{center}
We examine in detail various string scattering amplitudes in order
to extract the world-volume interactions of massless fields
on a Dirichlet brane. 
We find that the leading low-energy
interactions are consistent with the Born-Infeld
and Chern-Simons actions. In particular, our results confirm that the
background closed string fields appearing in these actions
must be treated as functionals of the {\it non-abelian} scalar fields
describing transverse fluctuations of the D-brane.
\vfill
\setcounter{page}{0}
\setcounter{footnote}{0}
\newpage

\section{Introduction} \label{intro}

Recent years have seen dramatic progress in the understanding
of nonperturbative aspects of string theory\cite{excite}.
With these studies has come the realization that extended
objects, other than just strings, play an essential role.
An important tool in these investigations has been the Dirichlet
brane\cite{joep}, which within the framework of perturbative
string theory provide an exact description of
of Ramond-Ramond charged solitons.

The world-volume theory of a single D-brane includes a massless
U(1) vector $A_a$, a set of massless scalars $X^i$, describing the transverse
oscillations of the brane, and their super-partner
fermions \cite{leigh}. The leading order low-energy action for these
fields corresponds to a dimensional reduction of a ten dimensional
U(1) super-Yang Mills theory. As usual in string theory, there are
higher order $\alpha'=\ls^2$ corrections,
where $\ls$ is the string length scale. As long as derivatives
of the field strengths (and second derivatives of the scalars)
are small compared to $\ls$, then the action takes a Born-Infeld
form \cite{bin}. This action also has a supersymmetric extension\cite{jhs}.
To take into account the couplings of the open string states
with closed strings, the Born-Infeld
 action can be extended to include background
closed string fields, in particular, the metric, dilaton and
Kalb-Ramond field. Thus one arrives at the following world-volume
action for a D$p$-brane:
\beq
S_{BI}=-T_p \int d^{p+1}\sig\ \Tr\left(e^{-\Phi}\sqrt{-det(\tG_{ab}+
\tB_{ab}+2\pi\ls^2\,F_{ab})}\right)
\labell{biact}
\eeq
where $T_p=(2\pi\ls)^{-p}/g$ is the brane tension with $g$
being the asymptotic closed
string coupling.\footnote{Our convention will be that the dilaton field $\Phi$
vanishes asymptotically far from the D-branes.}
Here, $F_{ab}$ is the field
strength of the world-volume gauge field, while
the metric and antisymmetric tensors are
the pull-backs of the bulk tensors to the D-brane world-volume, \eg
\beq
\tG_{ab}=G_{ab}+2G_{i(a}\,\prt_{b)}X^i+G_{ij}\prt_aX^i\prt_bX^j\ .
\labell{pull}
\eeq
The D-branes are also carry RR charge and so to account for the coupling
of the massless closed string RR states, the above action is supplemented
by a Chern-Simons action of the form\cite{mike,cs}
\beq
S_{CS}=\mu_p\int \Tr\left( e^{\tB+2\pi\ls^2\,F}\sum_n \tC_{(n+1)}\right)
\labell{csact}
\eeq
where $\tC_{(n+1)}$ are the pull-backs of the ($n$+1)-form RR potentials
and $\mu_p=\sqrt{2\pi}(2\pi\ls)^{3-p}$. Thus a D$p$-brane is naturally
charged under the ($p$+1)-form RR potential. However in the presence
of background Kalb-Ramond fields or world-volume gauge fields, it may
also carry a charge of the RR potentials with a lower form degree,
as allowed with the couplings induced by the exponential factor\cite{mike}.
Note that as well as the derivative couplings induced by the pull-back
as in eq.~\reef{pull}, in general the background fields would be functionals
of the scalars $X^i$ so that even in the leading low energy approximation
these fields would be governed by a non-linear sigma model type action.

One of the most remarkable aspects of D-brane physics is that the
U(1) gauge symmetry of an individual D-brane is enhanced to a
non-abelian U(N) symmetry for N coincident D-branes \cite{bound}.
 When N parallel
D-branes approach each other, the ground state modes of
strings stretching between the D-branes become massless. These
extra massless states carry the appropriate charges to then fill
out U(N) representations and the $U(1)^N$ of the individual D-branes
is enhanced to $U(N)$. Hence $A_a$ becomes a nonabelian gauge field
and the scalars $X^i$ becomes scalars in the adjoint representation
of U(N). Understanding how to accommodate this simple yet remarkable
modification in the world-volume actions for general backgrounds
is the primary motivation for the present paper, and we report some
progress in this direction. In trivial
backgrounds, the extension of the Chern-Simons action \reef{csact}
is apparently straightforward with the addition of the trace over
gauge indices of the non-abelian field strength which now appears in
the exponential factor. Because of its highly nonlinear nature,
the nonabelian extension of the Born-Infeld
action is unclear even for a flat space background. Tseytlin\cite{yet}
has made compelling arguments to suggest that one should supplement
the usual Born-Infeld form with a symmetrized trace over gauge
indices, \ie to leading order all commutators of the field strengths
should be dropped. Unfortunately this does not seem to capture the
full physics of the infrared limit\cite{notyet}, and it appears
nontrivial commutators must be included at sixth-order in the field
strength\cite{notyet2}. To go beyond trivial backgrounds is a question
which has received only limited attention\cite{moremike,extra}.
Douglas\cite{moremike} has proposed that the background fields
should be functionals of the non-abelian scalars (rather than, \eg only
the U(1) or center-of-mass component of $X^i$).
An additional point noted by Hull\cite{hull}
is that it is natural to assume that the pull-backs \reef{pull} in
both eqs.~\reef{biact} and \reef{csact}
are modified to be defined in terms of covariant derivatives of the full
adjoint scalar fields. That is $\prt_aX^i\rightarrow D_aX^i$ in
eq.~\reef{pull}.
As D-branes are objects originally defined within perturbative
string theory, these questions can be resolved by studying
string scattering amplitudes.
This is the approach of our present investigation, and we are
able to confirm proposals by Douglas and Hull. Thus the trace appearing
in the two pieces of the action above, eqs.~\reef{biact} and \reef{csact},
acts implicitly on not only the gauge fields but also on the
implicit gauge dependence in the background fields, both in the form
of the derivatives of the pull-backs
and the internal functional dependence.

The remainder of the paper is organized as follows: In section 2, we
examine amplitudes involving one closed string and one or two open
string states. These amplitudes allow us to identify interactions
linear and quadratic in the world-volume fields of the D-branes.
In section 3, we consider the world-volume interactions which
can be inferred from amplitudes for three open strings
and those for two closed strings. Note that in both of these
sections, we only consider massless bosonic modes of the string.
Finally, we present a brief discussion of our results in section 4.

Before continuing with our calculations, let us make a comment on
conventions. In the scattering amplitudes below, we will
set $\ls^2=\alpha'=2$. Our index conventions are that lowercase Greek
indices take values in the entire ten-dimensional
spacetime, \eg $\mu,\nu=0,1,\ldots,9$; early Latin indices take values
in the world-volume, \eg $a,b,c=0,1,\ldots,p$; and middle Latin indices
take values in the transverse space, \eg $i,j =p+1,\ldots,8,9$.
Thus, for example, $G_{\mu\nu}$ denotes the entire spacetime metric,
while $G_{ab}$ and $G_{ij}$ denote metric components for directions
parallel and orthogonal to the D-branes, respectively.

\section{Scattering Calculations} \label{scat}

In this section, we investigate the open-closed string couplings by
an analysis of the appropriate scattering amplitudes. We will consider
scattering amplitudes with one closed string and one open string in
the next subsection, and with one closed string and two open strings
in the subsequent subsection. The techniques for these calculations
(and in fact the amplitudes in the second case) are available in
refs.~\cite{scatc,ours,aki,scatd}, and we will refer the reader there for
details. One useful observation is that these amplitudes can all
be related to purely open string scattering amplitudes with
unusual kinematics, and so in fact one can determine the desired
amplitudes from much older calculations of open superstrings\cite{black}.
In fact, this result can be extended all tree-level open/closed string
amplitudes\cite{unpub}.

The results should be compared to those for the various interactions
arising in the actions \reef{biact} and \reef{csact}. So we begin
by calculating the interactions expected from these world-volume
actions. In particular, we expand
the actions for fluctuations around flat empty space, \ie
$G_{\mu\nu}=\eta_{\mu\nu}$, $B_{\mu\nu}=0=\Phi=C_{(n)}$. The fluctuations
should be normalized as the conventional field theory modes which
appear in the string vertex operators. As a first step, we transform
from the string frame to Einstein frame metric with
$G_{\mu\nu}=e^{\Phi/2}g_{\mu\nu}$. The latter only affects the
Born-Infeld action which becomes
\beq
S_{BI}=-T_p \int d^{p+1}\sigma\ \Tr\left(e^{{p-3\over4}\Phi}
\sqrt{-det(\tg_{ab}+e^{-\Phi/2}\tB_{ab}+2\pi\ls^2\,e^{-\Phi/2}F_{ab})}
\right)
\labell{biact2}
\eeq
Now with conventions of \cite{ours}, the string mode fluctuations take the
form
\beqa
g_{\mu\nu}&=&\eta_{\mu\nu}+2\kappa h_{\mu\nu}\nonumber\\
\Phi&=&\sqrt{2}\kappa\phi\nonumber\\
B_{\mu\nu}&=&-2\kappa b_{\mu\nu}\nonumber\\
C_{(n)}&=&4 c_{(n)}
\labell{normal}\nonumber
\eeqa
where in the present conventions $\kappa=2^3\pi^{7/2}\ls^4g$.
In the following, it is only the relative normalizations of these modes
which will be important.
Similarly the open string modes are normalized as
\beq
A_a={1\over\sqrt{T_p}2\pi\ls^2}a_a\qquad\qquad
X^i={1\over\sqrt{T_p}}\lambda^i
\labell{onormal}\nonumber
\eeq

Now continuing with the Born-Infeld action, it is straightforward,
although somewhat tedious, to expand eq.~\reef{biact2} using
\beqar
\sqrt{det(\delta^a{}_b+\M a b)}&=&
1+{1\over2}\M a a-{1\over4}\M a b \M b a+{1\over8}(\M a a )^2
\nonumber\\
&&\qquad\qquad
+{1\over6}\M a b \M b c \M c a-{1\over8}\M a a\M b c\M c b+
{1\over48}(\M a a)^3+\ldots
\eeqar
to produce a vast array of interactions. We are interested in
the interactions linear in the closed string fluctuations, and
linear or quadratic in the open string fields. 

We begin with the linear couplings of the closed strings to the
D-brane source itself (these will be useful in section 3, below)
\beq
\cL_0=-T_p\ka\left(h^a{}_a+{p-3\over2\sqrt{2}}\phi\right)
\labell{int0}
\eeq
Next there are interactions involving one closed string mode and
one open string mode
\beqa
\cL_0'&=&-\sqrt{T_p}\ka\,\Tr\left(\l^i\prt_ih^a{}_a+{p-3\over2\sqrt{2}}
\l^i\prt_i\phi\right)\nonumber\\
\cL_1&=&-\sqrt{T_p}\ka\,\Tr\left(2h_{ia}\prt^a\l^i-b_{ab}f^{ab}\right)
\labell{int1}
\eeqa
where $f_{ab}=\prt_aa_b-\prt_ba_a$. 
We have divided these interactions into two sets: those in $\cL_1$
appear naturally in the naive expansion where $f_{ab}$ appears explicitly
while $\prt_a\l^i$ appears in the pull-back of the metric \reef{pull}.
The interactions in $\cL_0'$ arise essentially from a Taylor expansion
in the transverse coordinates of the terms appearing in eq.~\reef{int0}.
Thus these terms arise because the transverse fluctuations of the D-branes
feel the variations of the background fields. Because only a single
field appears under the gauge trace, these interactions only involve the
U(1) or center-of-mass fluctuations. Finally we will need to compare
to the interactions involving a single closed string mode and two
open string modes
\beqa
\cL_0''&=&-{\ka\over2}\,\Tr\left(\l^i\l^j\prt_i\prt_jh^a{}_a+
{p-3\over2\sqrt{2}}\l^i\l^j\prt_i\prt_j\phi\right)\nonumber\\
\cL_1'&=&-\ka\,\Tr\left(2\l^j\prt_jh_{ia}\prt^a\l^i-
\l^j\prt_jb_{ab}f^{ab}\right)\nonumber\\
\cL_2&=&-\ka\,\Tr\left({1\over2}h^a{}_a(\prt\l)^2-h^{ab}\prt_a\l^i\prt_b\l_i
+h^{ij}\prt_a\l^i\prt^a\l^j+{p-3\over4\sqrt{2}}\phi(\prt\l)^2\right.
\nonumber\\
&&\qquad\qquad\left.\vphantom{{p-3\over2\sqrt{2}}}+{1\over4}h^a{}_a(f)^2
-h_{ab}f^{ac}f^b{}_c+{p-7\over8\sqrt{2}}\phi(f^2)-
2b_{ai}\prt_b\l^if^{ab}\right)
\labell{int2}
\eeqa
Here again the interactions in $\cL_0''$ and $\cL_1'$ arise from the
Taylor expansion of the previous interactions. At the level
of these quadratic interactions, we begin to include the non-Abelian
parts of the fluctuations. 

The expansion to determine the interactions from the Chern-Simons action
\reef{csact} is somewhat simpler. First we begin with the linear
couplings of the closed strings to the D-brane
\beq
\tL_0=4\mu_p{1\over(p+1)!}(c_{(p+1)})^{a_0\cdots a_p}
\epsilon^v_{a_0...a_p}
\labell{cint0}
\eeq
where $\epsilon^v_{a_0...a_p}$ is the volume-form on the D-brane.
Next we have interactions linear in both the RR fields and open
string modes
\beqa
\tL_0'&=&{4\mu_p\over\sqrt{T_p}}{1\over(p+1)!}\,\Tr\left(\l^i\prt_i
(c_{(p+1)})^{a_0\cdots a_p}\right)\,
\epsilon^v_{a_0...a_p}\nonumber\\
\tL_1&=&{4\mu_p\over\sqrt{T_p}}\,\Tr\left({1\over p!}\prt^{a_0}\l^i
(c_{(p+1)})_i{}^{a_1\cdots a_p}+{1\over2(p-1)!}f^{a_0a_1}
(c_{(p-1)})^{a_2\cdots a_p}\right)
\epsilon^v_{a_0...a_p}
\labell{cint1}
\eeqa
Finally we will also compare to the quadratic couplings
\beqa
\tL_0''&=&{4\mu_p\over{T_p}}{1\over2(p+1)!}\,\Tr\left(\l^i\l^j\prt_i
\prt_j(c_{(p+1)})^{a_0\cdots a_p}\right)
\epsilon^v_{a_0...a_p}\nonumber\\
\tL_1'&=&{4\mu_p\over{T_p}}\,\Tr\left({1\over p!}\l^i\prt^{a_0}\l^j\prt_i
(c_{(p+1)})_j{}^{a_1\cdots a_p}+{1\over2(p-1)!}f^{a_0a_1}
\l^i\prt_i(c_{(p-1)})^{a_2\cdots a_p}\right)
\epsilon^v_{a_0...a_p}\nonumber\\
\tL_2&=&{4\mu_p\over{T_p}}\,\Tr\left({1\over2(p-1)!}\prt^{a_0}\l^i
\prt^{a_1}\l^j
(c_{(p+1)})^{a_2\cdots a_p}+{1\over2(p-2)!}f^{a_0a_1}\prt^{a_2}\l^i
(c_{(p-1)})_i{}^{a_3\cdots a_p}\right.\nonumber\\
&&\quad\qquad\qquad\left.+{1\over8(p-3)!}f^{a_0a_1}f^{a_2a_3}
(c_{(p-3)})_i{}^{a_4\cdots a_p}
\right)\epsilon^v_{a_0...a_p}
\labell{cint2}
\eeqa
We have again divided the
interactions between those arising in the Taylor expansion of the background
(closed string) fields of previous
lower order terms and new terms coming from additional
factors of $f_{ab}$ or the pull-backs.

Note that certain interactions do not contribute at
leading order. For example, interaction $\Tr[B_{ab}F^{bc}F_c{}^a]$ does
not contribute in \reef{int2}, because the quadratic term
vanishes under the gauge trace.
However, this interaction will make contributions at higher orders
as the background field $B$ is Taylor expanded.

\subsection{Linear couplings} \label{one}

Here, we wish to compare the linear couplings in eqs.~\reef{int1}
and \reef{cint1} to the results of the appropriate scattering amplitudes.
As mentioned above, since these terms involve a single open string
state the gauge trace will select out only the U(1) component of
the open string fields -- below we will keep the Chan-Paton factors explicit
in any event. To compare to the Born-Infeld interactions \reef{int1},
we consider the scattering amplitude for 
one open NS and one closed NS-NS state. The
amplitude is given by
\beqa
A^{\rm NS,NS-NS}&\sim&\int\,dx_1\,d^2z_2\,\Tr<V^{\rm NS}(k_1,\z_1,x_1)
\,V^{\rm NS-NS}(p_2,\veps_2,z_2,\bar{z}_2)>
\labell{ans,nsns}\nonumber
\eeqa
The details of the vertex operators and the techniques in calculating
 may be found in refs.~\cite{scatc,ours,aki,scatd}.
The final result is
\beqa
A^{\rm NS,NS-NS}&\sim&i\left[2k_{1a}(\veps_2\cdot D)^{\mu
a}\z_{1\mu}-2k_{1a}(\veps_2\cdot D)^{a\mu}\z_{1\mu}-p_{2\mu}D^{\mu\nu}
\z_{1\nu}Tr(\veps_2\cdot D)\right]\,\Tr(T_1)
\nonumber
\eeqa
where $D^\mu{}_\nu$ is a diagonal matrix with values +1 on the world-volume
and --1 in the transverse space \cite{ours}.
Substituting the appropriate polarizations, one finds the following
scattering amplitudes for different states:
\beqa
A(\l,h)&\sim&2i(2k_{1a}\,\z_{1i}\,\veps_2^{ia}+p_2^i\,\z_{1i}\,
\veps_{2a}{}^a)\,\Tr(T_1)\labell{alh}\nonumber\\
A(\l,\phi)&\sim&i\frac{p-3}{\sqrt{2}}\,\z_{1i}\,p_2^i\, \Tr(T_1)
\labell{alp}\nonumber\\
A(a,b)&\sim&-2i(k_{1a}\,\z_{1b}-k_{1b}\,\z_{1a})\,\veps_2^{ab}\,\Tr(T_1)
\labell{aab}\nonumber
\eeqa
while $A(\l,b),\, A(a,h), \,A(a,\phi)=0$. Comparing these results to
eq.~\reef{int1}, it is clear that these string amplitudes are
precisely reproduced by a field theory calculation with those interactions.
The terms in $\cL_1$ reproduce $A(a,b)$ and the first term in $A(\l,h)$,
while those in $\cL_0'$ yield $A(\l,\phi)$ and the second term in
$A(\l,h)$. Meanwhile the vanishing of the remaining amplitudes is
consistent with the fact that there are no corresponding interactions
at this order.

To compare with the Chern-Simons interactions \reef{cint1} at this order,
we examine the
scattering amplitude between one NS open and one RR closed string
state, which is given by
\beqa
A^{\rm NS,RR}&\sim&\Tr\int\,dx_1\,d^2z_2\,<V^{\rm NS}(k_1,\z_1,x_1)
V^{\rm RR}(p_2,\veps_2,z_2,\bar{z}_2)>
\labell{ans,rr}
\eeqa
Again, we refer the reader to refs.~\cite{ours,aki} for the
details of the calculations, and simply state the final result
\beq
A^{\rm NS,RR}\sim\Tr(T_1)\z_{1\mu_0}\,
(F_{(n)})_{\mu_1...\mu_n}\,\epsilon^v_{a_0...a_p}\,\Tr[\ga^{\mu_0}\ga^{\mu_1}
...\ga^{\mu_n}\ga^{a_0}...\ga^{a_p}(1+\ga_{11})]
\labell{resrr}\nonumber
\eeq
where $F_{(n)}$ is the (Fourier transform of) the RR field strength
for the ($n-1$)-form potential $c_{(n-1)}$,
\ie $(F_{(n)})_{\mu_1...\mu_n}=i\,n\,p_{2[\mu_1}\veps_{2\,\mu_2...\mu_n]}$
-- see \cite{ours}.
Now choosing various explicit polarizations and
performing the trace over gamma matrices, one finds two nonvanishing
amplitudes:
\beqa
A(\l,c_{(p+1)})&\sim&\z_1^i\,(F_{(p+2)})_i{}^{a_0...a_p}\,
\epsilon^v_{a_0...a_p}\,\Tr(T_1)\nonumber\\
&\sim&i(\z_1^i\,p_{2i}\,\veps_{2}{}^{a_0...a_p}
+(p+1)\z_1^i\,k_1^{a_0}\,\veps_{2i}{}^{a_1...a_p})\,\epsilon^v_{a_0...a_p}
\Tr(T_1)\labell{alc}\nonumber\\
A(a,c_{(p-1)})&\sim&\z_1^{a_0}(F_{(p)})^{a_1...a_p}\,
\epsilon^v_{a_0...a_p}\,\Tr(T_1)\nonumber\\
&\sim&ip\,k_1^{a_0}\,\z_1^{a_1}\,\veps_2{}^{a_2...a_p}\,
\epsilon^v_{a_0...a_p}\,\Tr(T_1)\labell{aac}\nonumber
\eeqa
where we have used momentum conservation in the world-volume to rewrite
the amplitudes in both cases, \ie $k_1+p_2^\|=0$.
In this second form, it is clear that these amplitudes are reproduced
by the interactions given in eq.~\reef{cint1}. $\tL_1$ yields 
$A(a,c_{(p-1)})$ and the second term in $A(\l,c_{(p+1)})$, while
$\tL_0'$ corresponds to the first term in $A(\l,c_{(p+1)})$.

Hence in accord with expectations, it is clear that the background fields
in the world-volume actions \reef{biact} and \reef{csact} are functionals
of at least the U(1) or center-of-mass components of the scalar fields
$X^i$. This is necessary as the interactions arising from the Taylor
expansion of the closed string fields were necessary to properly reproduce
the string scattering amplitudes. A comment on these contributions to
the amplitudes is called for as for a strictly physical configuration
they will vanish. This arises because of the simple two particle kinematics.
Both particles are massless, \ie $k_1^2=0=p_2^2$, and combined with world-volume
momentum conservation \ie $k_1+p_2^\|=0$, this implies that $p_2^\bot=0$
as there are no null directions in the transverse space. Hence one only
really sees these interactions, in an analytic continuation of the
momentum in which one allows for $(p_2^\bot)^2=0$ without having
$p_2^\bot=0$.

A curious feature of the RR scattering amplitudes is that initially
they appear in a form involving the RR field strength, so that the
gauge invariance of these fields is manifest. However to match to the
expansion of the Chern-Simons action \reef{csact}, an integration
by parts is necessary and this gauge invariance is no longer manifest.

\subsection{Quadratic couplings} \label{quadi}

Next, we investigate the couplings of closed strings
quadratic in the open string fields. In fact, the appropriate
scattering amplitudes with two open strings and one closed string
have already been calculated in ref.~\cite{aki}, and we need only
interpret their results. Again, the reader may find the details of the
calculations there. One begins with an amplitude of the form
\beq
A\sim\int\,dx_1\,dx_2\,d^2z_3\,\Tr<V^{\rm NS}(k_1,\z_1,x_1)
\,V^{\rm NS}(k_2,\z_2,x_2)
\,V^{\rm closed}(p_3,\veps_3,z_3,\bar{z}_3)>
\labell{ampall}
\eeq
where the last vertex will either be a NS-NS or RR closed string vertex.
The calculation here may be related to that of a four point amplitude
of open superstrings, and the
final amplitude takes the form\cite{aki}
\beq
A\sim{\Ga[-2t]\over\Ga[1-t]^2}K(1,2,3)
\labell{finaki}
\eeq
where $t=-2k_1\cdot k_2$ and the kinematic factor may be obtained from
type I calculations\cite{black} with an appropriate substitution of
momenta and polarizations.
This amplitude has $t$-channel poles at
$t=\frac{1}{2},
\frac{3}{2},\frac{5}{2},...$ which correspond to propagation of on-shell
open string states.
Since we are only interested in the leading low energy
contact terms, we take the limit
$t\longrightarrow 0$ for which eq.~\reef{finaki} simplifies to
\beqa
A&\sim& -{1\over2t}K(1,2,3)\ .
\labell{acontact}\nonumber
\eeqa
There is actually no pole at $t=0$ in this expression as
the kinematic factor will provide a compensating factor of $t$.

Let us begin by considering the Born-Infeld couplings in detail, and
so we choose a NS-NS vertex in eq.~\reef{ampall}. Making the appropriate
explicit choices of polarizations, we find 
\beqa
A(\l,\l,h)&=&\nonumber\left(2k_1\cdot k_2\,\z_1\cdot\veps_3\cdot\z_2+k_1\cdot
k_2\,\z_1\cdot\z_2\,\veps_{3a}{}^a+\z_1\cdot p_3\,\z_2\cdot
p_3\,\veps_{3a}{}^a\right.\\
&&\nonumber\left.-2k_1\cdot\veps_3\cdot
k_2\,\z_1\cdot\z_2+4\z_1\cdot\veps_3\cdot k_1\,\z_2\cdot p_3
+(1\longleftrightarrow 2)\right)\Tr[T_1T_2] \\
\nonumber
A(\l,\l,\phi)&=&\frac{p-3}{2\sqrt{2}}\left(k_1\cdot
k_2\,\z_1\cdot\z_2+\z_1\cdot p_3\,\z_2\cdot p_3+(1\longleftrightarrow 2)\right)
\Tr[T_1T_2]
\\
\nonumber
A(\l,a,b)&=&-2i(2k_1^a\,\z_{1i}\,f_{2ab}\,\veps_3^{bi}-
\z_1\cdot p_3\,f_{2ab}\,\veps_3^{ab})\, \Tr[T_1T_2]\\
\nonumber
A(a,a,h)&=&2\left(\veps_{3ab}f_1{}^{ac}f_2{}^b{}_c
-{1\over4}f_{1ab}f_2{}^{ab}\,\veps_{3a}{}^a
+(1\longleftrightarrow 2)\right)\Tr[T_1T_2] \\
\nonumber
A(a,a,\phi)&=&-\frac{p-7}{4\sqrt{2}}(f_{1ab}f_2{}^{ab}
+(1\longleftrightarrow 2))\,\Tr[T_1T_2]
\labell{bosamp}
\eeqa
where in the last three amplitudes we have suggestively introduced
$f_{iab}=i(k_{ia}\z_{ib}-k_{ib}\z_{ia})$. One also finds that
$A(\l,\l,b)=A(\l,a,h)=A(\l,a,\phi)=A(a,a,b)=0$. It is not hard to verify
that these amplitudes are those precisely accounted for by the
interactions given in eqs.~\reef{int2}.

Next we consider scattering amplitude of two NS open and one RR closed
string states to compare with the quadratic interactions in the Chern-Simons
action. With a RR vertex, one finds that the amplitude \reef{ampall}
reduces to
\beqa
A^{\rm NS,NS,RR}&\sim&(F_{(n)}^3)_{\mu_1...\mu_n}\,\epsilon^v_{a_0...a_p}
\left(\Tr[\ga^{\mu_1}\ldots\ga^{\mu_n}\ga^{a_0}...\ga^{a_p}(1+\ga_{11})
\,\ga\cdot\z_2\,
\ga\cdot(k_1+p_3/2)\,\ga\cdot\z_1]\right.
\nonumber\\
&&\,\,\,+\Tr[\ga^{\mu_1}\ldots\ga^{\mu_n}\ga^{a_0}...\ga^{a_p}(1+\ga_{11})
\,\ga\cdot\z_1]k_1\cdot\z_2
\nonumber\\
&&\,\,\,-
\Tr[\ga^{\mu_1}\ldots\ga^{\mu_n}\ga^{a_0}...\ga^{a_p}(1+\ga_{11})
\,\ga\cdot\z_2]k_2\cdot\z_1
\nonumber\\
&&\left.\,\,-\Tr[\ga^{\mu_1}\ldots\ga^{\mu_n}\ga^{a_0}...\ga^{a_p}(1+\ga_{11})
\,\ga\cdot k_1]\z_2\cdot\z_1\right)\Tr(T_1T_2)
\labell{nsnsrr}\nonumber
\eeqa
Now it is straightforward to perform the trace over the Dirac matrices and find
$A^{\rm NS,NS,RR}$ in terms of only polarization and momenta. In
the end, one finds only three nonvanishing amplitudes
for a D$p$-brane, which may be expressed as
\beqa
A(\l,\l,c_{(p+1)})
&\sim&(\z_1\cdot p_{3}\,\z_2\cdot p_{3}\,\veps_{3}{}^{a_0...a_p}
+2(p+1)\,\z_1^i\,k_1^{a_0}\,\z_2\cdot p_3\,\veps_{3i}{}^{a_1...a_p}
\nonumber\\
&&\qquad\quad+
{p(p+1)}\,\z_1^i\,\z_2^j\,k_1^{a_0}\,k_2^{a_1}\,
\veps_{3ij}{}^{a_2...a_p})\,\epsilon^v_{a_0...a_p}
\Tr(T_1T_2)+(1\longleftrightarrow 2)\labell{allc}\nonumber\\
A(\l,a,c_{(p-1)})&\sim&i(\z_1\cdot p_3\,f_2^{a_0a_1}\,
\veps_3{}^{a_2...a_p}+(p-1)\z_1^i\,f_2^{a_0a_1}\,k_1^{a_2}\,
\veps_{3i}{}^{a_3...a_p})\,\epsilon^v_{a_0...a_p}
\Tr(T_1T_2)\labell{alac}\nonumber\\
A(a,a,c_{(p-3)})&\sim&-f_1{}^{a_0a_1}f_2{}^{a_2a_3}
\veps_3{}^{a_4...a_p}\,\epsilon^v_{a_0...a_p}
\Tr(T_1T_2)\labell{aaac}\nonumber
\eeqa
It is again straightforward to verify that these are the amplitudes
precisely accounted for by the interactions in eq.~\reef{cint2}.

\section{Additional Amplitudes} \label{prelim}

In this section, we consider two classes of amplitudes which do not address
directly the question of interactions between open and closed string states.
However, they do add to the coherent picture of consistency between the
string scattering amplitudes and the effective actions discussed in
the introduction. The amplitudes considered here are for (i) three
open strings and (ii) two closed strings. 

\noindent {\it Three massless open strings:} Here there are various
distinct
amplitudes depending on the choices for the polarizations of the
particles. For three vectors (\ie all polarizations parallel to the
D-brane world-volume), the amplitude takes the form
\beq
(\z_1\cdot\z_2\,\z_3\cdot
k_1+\z_2\cdot\z_3\,\z_1\cdot k_2+\z_1\cdot\z_3\,\z_2\cdot k_3)
\Tr( T_1 [T_2,T_3])
\labell{vect}
\eeq
where the commutator of the U(N) Chan-Paton factors arises from including
the two distinct cyclic orderings of the vertex operators. This result
is of course identical to that in ten-dimensional open string theory
(D9-branes backgrounds) which was calculated long ago, and was known
to reproduce the three-point interaction for three non-Abelian gauge
particles arising in the minimal action $\Tr(F_{ab}F^{ab})$
\cite{ymill,symill}. This is consistent then with an expansion to leading
order of the Born-Infeld action \reef{biact}. Also of interest is the
observation that the superstring scattering amplitude\cite{symill}
does not contain
any terms of order $k^3$, and so the low energy action does not contain
a cubic term of the form $\Tr(F_{ab}F^{bc}F_c{}^a)$. The absence of this
term is also consistent with the expansion around
flat space of the Born-Infeld action
\reef{biact} accompanied with a symmetrized trace prescription for
the gauge indices. The latter observation was also recently made from
an analysis of the beta functions\cite{perry}. Such terms do arise for
the bosonic string, and can be accommodated by including a Born-Infeld
action with an antisymmetric trace prescription \cite{phil}.
In fact for the superstring, one could extend this analysis to
consider four-point functions which would reveal an $F^4$ contact
term\cite{yeti}, which is in fact again consistent with the expansion
of Born-Infeld with a symmetrized trace. It is known that certain
inconsistencies arise, however, between string theory and this
low energy action at order $F^6$ \cite{notyet,notyet2}.

The other nonvanishing amplitude involves two scalars (\ie polarizations
orthogonal to the D-brane world-volume) and one vector. Although the
details of the scalar vertex operators differ, the resulting amplitude
is essentially the same as eq.~\reef{vect} above, with the restriction
that $\z_{1,2}$ are orthogonal to all momentum vectors (which lie
in the D-brane world-volume). Hence, one finds
\beq
\z_1\cdot\z_2\,\z_3\cdot k_1\,\Tr(T_1[T_2,T_3])\ .
\labell{scalar}\nonumber
\eeq
This result reproduces the appropriate non-Abelian gauge
interaction occurring in the scalar action $\Tr(D_a X^i D^a X^i)$,
which of course arises from a dimensional reduction of $F^2$ in
ten dimensions. This again consistent with the expansion of the
Born-Infeld action around flat space, as long as the pull-backs
are extended to involve gauge covariant derivatives of the
non-abelian scalar fields, \eg eq.~\reef{pull} is replaced by
\beq
\tG_{ab}=G_{ab}+2G_{i(a}\,D_{b)}X^i+G_{ij}D_aX^iD_bX^j\ .
\labell{pullg}
\eeq
Hull \cite{hull} noted that this modification is a natural choice
which accommodates the non-Abelian gauge symmetry.

Finally one may observe that the vanishing of the scattering amplitudes
with three scalars or two vectors and a scalar is also
consistent with the leading order expansion of the Born-Infeld
action \reef{biact}. One can also check the consistency of the
four-point amplitudes involving scalars with the expansion of the
Born-Infeld action\cite{joebook}, and of course consistency
arises from the dimensional reduction of the $F^4$ terms in ten
dimensions which are known to agree\cite{yet}.

\noindent {\it Two closed strings:} The scattering amplitudes
for scattering two massless closed strings from a D-brane were
presented in refs.~\cite{scatc,ours}. In ref.~\cite{ours}, it was
verified that the scattering was consistent with the propagation
of the closed string in the long range fields of a supergravity
solution corresponding to a D-brane.  Essentially, this analysis
showed that the linear coupling
of the closed string fields agreed with that appearing in the low-energy
world-volume action (\ref{biact},\ref{csact}). For example, the
Born-Infeld term gave in eq.~\reef{int0}
\beq
\cL_0=-T_p\ka\left(h^a{}_a+{p-3\over2\sqrt{2}}\phi\right)
\labell{int00}\nonumber
\eeq
Thus the linear
coupling of antisymmetric Kalb-Ramond field $B^{\mu\nu}$ is zero.
Similarly the linear source term for the RR potentials agrees
with that from the Chern-Simons action \reef{csact}, as given above
in eq.~\reef{cint0}
\beq
\tL_0=4\mu_p{1\over(p+1)!}(c_{(p+1)})^{a_0\cdots a_p}
\epsilon^v_{a_0...a_p}
\labell{cint00}\nonumber
\eeq
This analysis of scattering amplitudes
can be extended to the case of D-branes with a
constant background gauge field or equivalently $B$ potential\cite{mrg}.
These backgrounds modify the worldsheet boundary conditions and the
above source terms. For example, a linear coupling to
the antisymmetric $B$-field appears.

A more careful analysis of these amplitudes can also determine
the quadratic contact terms for the closed strings themselves
on the D-brane world-volume. In the small momentum limit,
\ie $\ls\rightarrow
0$, one finds massless poles in both the $s$- and $t$-channels. The
latter, which were considered in the previous analysis,
result from the absorption of a massless closed string
by the D-brane. In the $s$-channel, these poles arise because the 
closed strings can couple
to a single massless open string through the interactions
in eqs.~\reef{int1}
and \reef{cint1}, which then can propagate on-shell (or nearly on-shell)
along the world-volume. Subtracting the field theory amplitudes in
both of these channels from the string
amplitudes leaves a set of quadratic contact terms for the closed
strings on
the world-volume. To leading order in the derivative expansion, there
are non-derivative couplings, which we will extract below. 

As before, let us begin by determining the interactions
which an expansion of the actions \reef{biact} (or rather \reef{biact2})
and \reef{csact} predict. First from the Born-Infeld action, one
finds\footnote{Throughout the following, we ignore the non-Abelian
trace, which would have the trivial effect of introducing an overall
factor of $N$ in the amplitudes and interactions.}
\beq
\cL_{0,2}=-T_p\ka^2\left(\frac{1}{2}(h^a{}_a)^2-h_{ab}h^{ab}
+b_{ab}b^{ab}+\left({p-3\over4}\right)^2\phi^2
+{p-3\over2\sqrt{2}}\phi h^a{}_a\right)
\labell{int02}
\eeq
while the Chern-Simons action yields
\beq
\tL_{0,2}=-8\kappa\mu_p{1\over2(p-1)!}b^{a_0a_1}(c_{(p-1)})^{a_2\cdots a_p}
\epsilon^v_{a_0...a_p}
\labell{cint02}
\eeq

Next consider the amplitudes for two closed string states scattering
off of a D-brane, which were calculated in \cite{ours}. Expanding the
amplitudes around small $t=-k^2$ and $s=-q^2$, one finds\footnote{Note
that there is a change in conventions between \cite{ours} and the
present paper: $(T_p)_{\rm there}=(\ka T_p)_{\rm here}$.}
\beq
A=-i\ka^2T_p\left(\frac{a_1}{k^2}-\frac{a_2}{4q^2}
+\ldots\right)
\labell{astring}
\eeq
where $a_1$ and $a_2$ are expressions involving the closed string
polarizations and are quadratic in momenta --- see \cite{ours}.
An important point in the following is that all of the higher
order terms which are not explicitly shown above are at least quadratic
in the momenta.

To extract the desired contact terms, one now needs to calculate the
corresponding field theory scattering amplitudes, in $s$-channel using
the action \reef{biact}, and $t$-channel using bulk supergravity action.
We will present some of the details for the calculation for scattering with a
graviton and a dilaton. For this case, the field theory $t$-channel amplitude
appears in \cite{ours}:
\beqa
A'_t(h,\phi)&=&i\ka^2 T_p{p-3\over\sqrt{2}}{1\over k^2}p_2\cdot\veps_2
\cdot k
\labell{tchan}
\eeqa
Now the $s$-channel amplitude can be calculated as:
\beqa
A'_s(h,\phi)&=&\tilde{V}^i_{h\l}\,\tilde{G}^{ij}_{\l}\,\tilde{V}^j_{\phi
\l} \nonumber
\eeqa
where $\tilde{G}^{ij}_{\l} =-iN^{ij}/q^2$ is the scalar propagator
on the D-brane world-volume, and the vertex functions are derived
for the linear interactions in eq.~\reef{int1}
\beqa
\tilde{V}^i_{h\l}&=&\ka\sqrt{T_p}(\veps_{1a}{}^a\,p_1^i
-2p_1^a\,\veps_{1a}{}^i) \nonumber\\
\tilde{V}^i_{\phi\l}&=&\ka\sqrt{T_p}{p-3\over2\sqrt{2}}\,p_2^i
\nonumber
\eeqa
Substituting these into  field theory scattering amplitude above, 
 one will find
\beqa
A'_s(h,\phi)&=&-i\ka^2{T_p}{p-3\over2\sqrt{2}}{1\over q^2}
\left(\veps_{1a}{}^a\,p_1\cdot N\cdot p_2+2p_1\cdot N\cdot
\veps_1\cdot N\cdot p_2\right)
\labell{schan}
\eeqa
Now subtracting these field theory amplitudes (\ref{tchan},\ref{schan})
from the corresponding string amplitude \reef{astring}, one finds
\beqa
A(h,\phi)-A'_t(h,\phi)-A'_s(h,\phi)&=&-i\ka^2T_p{p-3\over2\sqrt{2}}
\veps_{1a}{}^a+\ldots
\labell{contamp}
\eeqa
Again, the important point to note is that the terms implicitly
denoted by the ellipsis are all at least quadratic in momenta.
Hence the world-volume theory must include a non-derivative interaction
between the graviton and dilaton field. In fact, it is not hard to
see that the above amplitude \reef{contamp} is precisely reproduced
by the final term in eq.~\reef{int02}.

Similar calculations for the other massless closed string modes
yield
\beqa
A(h,h)-A'_t(h,h)-A'_s(h,h)&=&-i\ka^2T_p\left(\veps_{1a}{}^a\veps_{2b}{}^b
-2Tr(\veps_1\cdot V\cdot\veps_2\cdot V)\right)+\ldots
\nonumber\\
A(b,b)-A'_t(b,b)-A'_s(b,b)&=&2i\ka^2T_pTr(\veps_1\cdot V\cdot\veps_2\cdot
V)+\ldots
\nonumber\\
A(\phi,\phi)-A'_t(\phi,\phi)-A'_s(\phi,\phi)&=&
-2i\ka^2T_p\left(\frac{p-3}{4}\right)^2+\ldots
\labell{contampb}\nonumber
\eeqa
These amplitudes are precisely reproduced by the remaining terms in
eq.~\reef{int02}. One also finds that the string amplitudes
$A(b,\phi)$ and $A(h,b)$ vanish, as well as the corresponding
field theory amplitudes. Hence no quadratic contact terms appear
in these cases, which is again consistent with the absence of
such interactions in eq.~\reef{int02}.

Similar calculations for scattering amplitudes involving the massless
RR states yield quadratic contact terms consistent with the interactions
in eq.~\reef{cint02}. That is the only non-derivative contact terms
appear for the combination of $b$ and $c_{(p-1)}$.

\section{Discussion} \label{discuss}

By directly examining various string scattering amplitudes, we have 
extracted various world-volume interactions for the effective
field theory on D-branes. Our results are entirely consistent
with the Born-Infeld and Chern-Simons actions (\ref{biact},\ref{csact}),
as long as these expressions are properly interpreted. Our analysis
suggests that the pull-backs of the background field tensors
must be constructed using gauge covariant derivatives of the
non-abelian scalar fields, as in \reef{pullg} for example.
Also the background closed string fields in both expressions are
functionals of these non-abelian scalars. The latter conclusion
is evident from the analysis of the quadratic couplings considered
in section 2.2. Thus the gauge trace appearing in eqs.~\reef{biact}
and \reef{csact} applies not only on the gauge fields, but also on the
non-abelian scalars in the background fields, both in the form
of the covariant derivatives of the pull-backs
and the internal functional dependence. Our scattering amplitude
analysis can be extended to the case of mixed boundary conditions,
as in \cite{mrg}. Such boundary conditions represent D-branes with
a constant U(1) gauge field or equivalently $B$ potential, and so
extra interactions are revealed in the scattering amplitudes.
Although we have not pursued these calculations in the same detail
as those presented above, the results again seem to be entirely
consistent with the structure of the Born-Infeld and Chern-Simons
actions.

Given the above results, consider the case, for example, of a
(test) D-brane propagating in a curved (fixed) background geometry
with metric $G^0_{\mu\nu}(x^\rho)$. To implement the necessary functional
dependence in \reef{biact}, we can adapt the spacetime coordinates for
a static gauge choice, at least in
a neighborhood of the D-brane world-volume, \ie we make a split of
world-volume coordinates, $x^a=\sigma^a$, and transverse coordinates,
$x^i={1\over N}\Tr\,X^i$. Then the metric functional appearing in
the D-brane action would be given a {\it non-abelian} Taylor expansion
\beq
G_{\mu\nu}=\exp\left[X^i{\prt\ \over\prt {x^i}}\right]G^0_{\mu\nu}
(\sigma^a,x^i)|_{x^i=0}
\labell{taylor}
\eeq
Similar expansions would also be applied for other nontrivial
background fields.
Such a Taylor expansion naturally produces an expression symmetric in
the $X^i$ since the partial derivatives commute. Our analysis,
however, leaves open many questions of ordering the various fields
under the gauge trace. For example,
how are the $X^i$'s in eq.~\reef{taylor} ordered
between various components
of the background fields appearing in higher order interactions,
or with field strength components of the world-volume gauge field.

The latter might be included in the question of finding the complete
definition for the gauge trace for the Born-Infeld action \reef{biact}.
However, new ambiguities are also appearing here for the
Chern-Simons action \reef{csact} as well.
Essentially, our analysis does not resolve these questions
because we cannot see any nontrivial commutator terms with the present
amplitudes. In principle, our method can be extended to higher
point amplitudes to begin to address these ambiguities, however,
in practice, extracting the higher order contact interactions
would be extremely tedious. We have made some progress in examining
scattering amplitudes involving one closed and three open strings \cite{prep}.
It is evident from other investigations\cite{moremike,notyet2}
that commutator terms play an important role at higher orders.

Recall that in the amplitude $A^{\rm NS,RR}$ in eq.~\reef{ans,rr},
one starts with a vertex operator written in terms of the RR
field strength. Hence the resulting interactions are naturally
derived in terms of this field strength, and as a result are
invariant under the RR gauge transformations. However,
one must integrate by parts and thus the gauge
invariance is no longer manifest. Consider the linear interactions
involving $c_{(p+1)}$ in eq.~\reef{cint1}. Invariance
under gauge transformations
depending only on world-volume directions is clear, \ie a total
derivative on the world-volume is produced for the term
in $\tL_1$ which originates in the pull-back. However for
transformations depending on transverse directions\footnote{Note
that here we would consider the gauge parameters as functionals of the
non-abelian scalars, just as above for the background fields.}, this
term alone
is not invariant, but invariance results from a cancellation with
a contribution from the term in $\tL_0'$ which originates in the
Taylor expansion of the background. So there is an important interplay
of these two terms, although they have completely different origin
in the Chern-Simons action. Similar comments apply for higher order
interactions, as well.

A similar but more complicated discussion would apply for the
interactions in the Born-Infeld action \reef{biact}. For example,
while the non-abelian Taylor expansion presented in eq.~\reef{taylor}
is completely non-covariant, the relevant string scattering amplitudes
will be covariant. It may be that gauge invariance can provide a 
tool to resolve some of the ordering questions raised above.
It remains a problem, however, that in general the static gauge choice
made to construct the Taylor expansion \reef{taylor}
will only apply in a local neighborhood
of the D-brane world-volume. It seems that the description of the
world-volume dynamics of D-branes is still lacking at some basic level.
The disparity between the role of world-volume and transverse coordinates
seems a fundamental problem. Resolving this disparity may provide
further insight into the non-abelian nature of spacetime or general
backgrounds as seen by string theory.

As a final note, we will comment on the physical significance of
some of the interactions considered in this paper. To focus the
discussion, consider the interaction $\eps^{abc}\Tr[f_{ab}\l^i]
F_{(2)ci}$ which appears in the Chern-Simons action of a D2-brane.
Here,  $F_{(2)}$ is the field strength of RR one-form potential.
As discussed above,
an integration by parts has been performed to combine two terms
from $\tL'_1$ and $\tL_2$ in eq.~\reef{cint2} to produce this
covariant expression. To understand the physical role played
by this interaction, consider two D2-branes which coincide to
produce a  $U(2)=U(1)\times SU(2)$ gauge symmetry on the world-volume.
Now a constant background field strength $f_{ab}$
in the $U(1)$ will induce a coupling
to the one-form RR potential through the higher order interactions
in the Chern-Simons action \cite{mike}. The resulting configuration
can be thought of as a bound state of D2- and D0-branes \cite{joep,jake}.
Giving an expectation value to a $SU(2)$ field strength in, \eg
the $\sigma_3$ part of the gauge group, yields no such coupling since
$Tr f_{ab}=0$. This indicates there
is no net D0-brane charge in the system. One can think of the individual
eigenvalues of the gauge matrix as indicating that the two fundamental
D2-branes carry equal and opposite densities of D0-branes, and hence
the net charge is zero. If now, however, one gives an expectation 
value to some $\l^i$ in the $\sigma_3$ part of the gauge group, one
is separating one D2-brane from the other.
While there is still
no net charge, this produces a small separation of
positive and negative charges, \ie a dipole, which should be
detectable in scattering closed strings. The coupling
above, $\eps^{abc}\Tr[f_{ab}\l^i]F_{(2)ci}$, provides precisely the
desired dipole coupling for the RR one-form. A similar discussion
extends to the analogous couplings on other D$p$-branes. As well,
higher order interactions may play the role of higher multi-pole
couplings for the appropriate configurations of world-volume fields.
These results may be of interest in examining various nontrivial
nonabelian field configurations which have been discussed recently
\cite{shape}.

\vspace{1cm}
{\bf Acknowledgements}

RCM was supported by NSERC of Canada, Fonds FCAR du Qu\'ebec
and at the ITP by NSF Grant PHY94-07194. RCM would also like
to thank the Aspen Center for Physics for hospitality while this
paper was finished.
We would also like to acknowledge useful conversations with
F. Ardalan, H. Arfaei and A. Hashimoto.

\end{document}